\documentclass[12pt]{article}
\usepackage{psfig}
\newcommand{\beq}{\begin{equation}}
\newcommand{\eeq}{\end{equation}}
\def\bea{\begin{eqnarray}}
\def\eea{\end{eqnarray}}
\newcommand{\ba}{\begin{array}}
\newcommand{\ea}{\end{array}}

\newcommand{\gsim}{ \mathop{}_{\textstyle \sim}^{\textstyle >} }
\newcommand{\lsim}{ \mathop{}_{\textstyle \sim}^{\textstyle <} }

\def\rc{r_c}
\def\pikrc{\pi{}k\rc}
\def\z1{e^{\pikrc}}

\def\planck5d{M_{{\rm 5d}}}
\def\gauge5d{g_{{\rm 5d}}}
\def\yukawa5d{y_{{\rm 5d}}}
\def\quartic5d{\lambda_{{\rm 5d}}}
\def\L{{\rm L}}
\def\R{{\rm R}}
\def\B{{\rm B}}
\def\D{{\rm D}}
\def\PsiL#1{\Psi_{\L, \B }^{(#1)}}
\def\PsiR#1{\Psi_{\R, \B }^{(#1)}}
\def\psileft#1#2{\PsiL{#1}\!\left(#2\right)}
\def\psiright#1#2{\PsiR{#1}\!\left(#2\right)}
\def\leftmode#1#2{\xi_{#1}\!\left(#2\right)}
\def\rightmode#1#2{\eta_{#1}\!\left(#2\right)}
\def\gaugemode#1#2{{ \chi}_{#1}\!\left(#2\right)}

\def\gaugemass#1{M_{{#1}}}
\def\eigen#1{\lambda_{#1}}

\begin{document}

\title{\vskip-2.5truecm{\hfill \baselineskip 14pt {{
\small  \\    
\hfill FTUV-01-0111 \\
\hfill IFIC/01-1 \\
\hfill March 01}}\vskip .9truecm}
 {\bf Dynamical Symmetry Breaking in Warped Compactifications  }} 

\setcounter{footnote}{0}

\vspace{5cm}

\author{ N. Rius\footnote{nuria@tlaloc.ific.uv.es} \, and V. Sanz\footnote{sanz@tlaloc.ific.uv.es}
 \\  \  \\
{\it Depto. de F\'\i sica Te\'orica and IFIC }\\
{\it Universidad de Valencia-CSIC, Valencia, Spain}\\
}
\date{}
\maketitle
\vfill

\begin{abstract}
\baselineskip 20pt
We study dynamical electroweak symmetry breaking in the Randall-Sundrum 
scenario. We show that one extra dimension is enough to give the correct 
pattern of electroweak symmetry breaking in a simple  
model with gauge bosons and the right-handed top quark in the bulk.
The top quark mass is also in agreement with experiment. 
Furthermore, we propose an extended scenario with all Standard Model 
gauge bosons and fermions propagating in the bulk, which naturally 
accommodates the fermion mass hierarchies.  
No new fields or interactions beyond the observed in the Standard Model 
are required.

\end{abstract}

\vfill
\thispagestyle{empty}

\newpage
\pagestyle{plain}
\setcounter{page}{1}

\renewcommand{\thefootnote}{\arabic{footnote}}
\setcounter{footnote}{0}

\section{Introduction}

The origin of the electroweak symmetry breaking is one of the most important 
questions in particle physics. In the Standard Model (SM) it is achieved 
by a nonzero vacuum expectation value of a fundamental scalar Higgs field.
However, the squared-mass of a fundamental scalar field receives 
quadratically divergent radiative corrections, hence suffers from the 
so-called hierarchy problem if the cutoff scale is much higher than the 
electroweak scale. This is indeed the case in the SM, where a large  
hierarchy exists between the electroweak scale ($\sim 100$ GeV) and 
the Planck scale ($M_P \sim 10^{18}$ GeV).
 
It has been recently realized that this hierarchy of scales could have 
its origin in the presence of extra dimensions with nontrivial 
space-time geometries\cite{XL,RS}.
If we live in $D=4+\delta$ dimensions, there 
is the possibility that the Planck scale $M_P$ is 
actually an effective four dimensional scale determined by the fundamental 
scale of the $(4+\delta)$-theory, $M$, and the geometry of space-time. 
An explicit example of this is the Randall and Sundrum model \cite{RS},
where the hierarchy problem is solved by introducing a 
warped extra dimension. 
The space-time is a slice of $AdS_5$ with one extra dimension compactified 
on an orbifold, $S^1/Z_2$, of radius $r_c$.
The metric is given by 
\begin{equation}
ds^2=e^{-2k|y|}\eta_{\mu \nu}dx^{\mu}dx^{\nu}+dy^2
\label{RSmetric}
\end{equation}
where $y$ is the fifth coordinate, $0 \le y \le \pi r_c$, 
$\mu$, $\nu$ are four dimensional indices, 
$k$ is the $AdS_5$ curvature of order of the Planck scale, and 
$e^{-2k|y|}$ is called the warp factor.
The Randall-Sundrum scenario consists of two three-branes located at the 
orbifold fix points $y_{*}=0,\pi r_c$. From (\ref{RSmetric}) one can see that 
the warp factor determines the physical energy scale at the position $y$ from 
the point of view of a 4$\D$ observer. 
Thus assuming $k r_c \sim 12$, the physical scale of the brane located at 
$y_{*}=\pi r_c$ is given by $k e^{-\pi k r_c} \sim (100-1000)$ GeV. 
This sets the scale on the brane to be the electroweak scale and solves 
the hierarchy problem.

The hierarchy problem can also be avoided if the Higgs is a composite 
object rather than a fundamental field, and ceases to be a dynamical degree 
of freedom not much above the electroweak scale. 
Since the top quark  is the heavier fermion in the standard model, it has 
been the first candidate to form the composite Higgs, bound out of the 
third generation weak doublet
$\psi_L$ and the right-handed top field, $t_R$.
Such composite Higgs arises naturally in the presence of some strongly 
coupled four-quark operators.
In four dimensional top condensate models, though, there is a problem
when one tries to accommodate the top quark mass in the experimental 
range as well as electroweak symmetry breaking at the correct scale, 
because the large top Yukawa 
coupling gives a top quark too heavy \cite{MTY},\cite{BHL}. 
This problem can be solved if a
new vector-like fermion is introduced with the same quantum numbers of 
$t_R$, which becomes the appropriate constituent of the 
Higgs boson together with $\psi_L$ \cite{seesaw}.
While this mechanism neatly accommodates both the measured top quark mass 
and a Higgs vev of the electroweak scale, the drawback is that
one has to include additional structures to originate the 
non-renormalizable four-quark interactions.
  
In this work we will show that in the Randall-Sundrum scenario it is 
possible to construct a phenomenologically successful effective theory 
with no fundamental scalars, including just the symmetry group 
$SU(3)_C \times SU(2)_W \times U(1)_Y$ and 
the SM fermion and gauge boson fields.
Dynamical electroweak symmetry breaking in the presence of flat extra 
dimensions has already been considered in the literature
\cite{AD}-\cite{D2}. In particular, 
it has been noticed that the ingredients needed 
for dynamical electroweak symmetry breaking are naturally present in the SM,
provided the gauge bosons and some fermions propagate in extra dimensions 
compactified at the $\sim$ TeV scale \cite{D1,D2}.   
Four-quark operators are always induced by QCD in compact dimensions, 
via the exchange of the Kaluza-Klein (KK) excitations of the 
gluons.
Moreover, by allowing the right-handed top quark to live in the bulk, 
its KK modes will naturally play the role of the required 
vector-like quark.

The mechanism outlined above does not work in $D=5$ flat space because the
interchange of gluon KK excitations is not strong enough to  
form the quark condensate. 
It is necessary to assume a higher dimensionality
($D \ge 6$), thus the degeneracy of the gluon KK modes makes the 
relevant four-quark operator stronger.
However, in the Randall-Sundrum scenario the coupling between KK gauge 
bosons and fermions (both living in the bulk and on the TeV brane)
can be larger than in the flat space case \cite{KK1},
so one extra dimension is enough to trigger dynamical 
electroweak symmetry breaking
without introducing fundamental scalar fields.

In section 2 we present a minimal set-up which breaks correctly the
electroweak symmetry, 
and we briefly review the KK decomposition of massless gauge boson and 
fermion fields in the Randall-Sundrum scenario. 
In the next section, we  use these results to construct an effective theory 
with four-fermion operators, and we argue that the binding strength of such 
operators is large enough to form bound states, one of which will
be identified as the Higgs field.  
We compute the effective scalar Lagrangian in section 4.
Finally, we discuss fermion masses both in the simplest model 
(section 5) and in an alternative scenario which  
provides an explanation of the observed fermion mass hierarchies (section 6).
We conclude in section 7.

\section{The simplest set-up}
\label{setup}

In the Randall-Sundrum model \cite{RS} only gravity propagates in 
the 5D bulk, while the SM fields are confined on the TeV brane.
Subsequently, the phenomenological consequences of placing the SM fields 
in the bulk have been extensively studied \cite{KK0}-\cite{AS}.

In this section, we describe a minimal set-up which leads to 
dynamical electroweak symmetry breaking without the need for a 
fundamental Higgs field. 
We begin by studying a toy model with one generation of fermions, 
the third one, and postpone the discussion of flavor symmetry 
breaking to section \ref{fm}.
We consider that gluons live in the 5D bulk, so their KK modes
strongly coupled to quarks can induce the formation of 
bound states. 
As we will see in section \ref{fm}, in order to obtain the correct 
value of the top mass also the right-handed top quark  
should propagate in 5D. 
For simplicity, we assume that the remaining third generation fermions are 
confined on the TeV brane, $y_{*}=\pi r_c$.

Since the right-handed top carries hypercharge, the $U(1)_Y$ gauge boson 
propagates in the bulk, while the $SU(2)_W$ gauge bosons can either reside 
on the TeV brane or propagate in the bulk, 
because we do not require the components of weak doublet 
fermions  to be in different places along the fifth dimension.
For definiteness we consider that the $SU(2)_W$ gauge bosons also live 
in 5D, but our conclusions are completely
independent of this assumption.

We are going to study the dynamical generation of masses through the 
condensation of a pair quark-antiquark, so we shall use the KK decomposition 
of 5$\D$ massless fields in the Randall-Sundrum model.

\subsection{Fermion field}

Consider a 5$\D$ massless fermion field  $\Psi(x,y)$. 
Compactifying on an orbifold $S^1/ Z_2$ we can choose the zero mode to be a 
left- or right-handed fermion. Imposing that the bulk fermion is even under 
this compactification,

\begin{equation}
\gamma_5 \ \Psi(x,-y) = +  \Psi(x,y) \ ,
\label{bcf}
\end{equation}
only the right-handed  zero mode survives. We will identify this zero mode 
with the right-handed top quark $t_R$. The Kaluza Klein decomposition for a 
bulk fermion with boundary conditions (\ref{bcf}) can be written as 
\cite{KK3} 
\begin{eqnarray}
\Psi(x,y) = \sum_n \left[\psileft{n}{x} \leftmode{n}{y}+ \psiright{n}{x}\rightmode{n}{y}\right] 
\label{fer1}
\end{eqnarray}
where
\begin{eqnarray}
\leftmode{n}{y}
 &=& \sqrt{\frac{2 k}{1-e^{-\pi{}k\rc}}}\,
     e^{-\frac{k}{2}|\pi\rc-y|} e^{\frac{3}{2} k |y|}
     \sin\left\{\frac{m_n}{k}\left(e^{k |y|}-1\right) \right\} \ ,
\nonumber\\
\rightmode{n}{y}
 &=& \sqrt{\frac{2 k}{1-e^{-\pi{}k\rc}}}\,
     e^{-\frac{k}{2}|\pi\rc-y|} e^{\frac{3}{2} k |y|}
     \cos \left\{ \frac{m_n}{k}\left(e^{k |y|}-1\right) \right\} \ ,
\label{fkk}
\end{eqnarray}
and  $m_n = n \pi k/(e^{\pikrc}-1)\neq{}0$. For the zero mode 
\begin{equation}
\leftmode{0}{y}  = 0, \qquad
\rightmode{0}{y} = \sqrt{\frac{k}{1-e^{-\pi{}k\rc}}}\,
                     e^{-\frac{k}{2}|\pi\rc-y|} e^{\frac{3}{2} k |y|} \ .
\label{zeromode}
\end{equation}
Note that massless bulk fermions are localized near the TeV brane, 
$y_*= \pi r_c$, due to the factor $e^{-k|\pi r_c -y|}$ of the wave 
function for all modes. 

A 5D Lorentz invariant gauge theory has no chiral anomalies because the
fermion representation is vector-like. However, the boundary conditions 
imposed above prevent the existence of the $\Psi_L$ zero-mode and
 we have to worry about anomalies in the bulk. 
This problem can be solved by including a  Chern-Simons term in the 
action,  which makes the scenario presented in this 
paper anomaly-free \cite{D1}.

\subsection{Gauge bosons}

The bulk gauge bosons in the
Randall-Sundrum scenario have been discussed in \cite{KK1}.  
We work in the gauge $\partial^\mu A_\mu = 0$ and 
$A_5=0$,  with orbifold conditions 
\begin{eqnarray}
\partial_5{}A_\mu\!\left(x,y\!=\!y_{*}\right)
\,=\,0\,=\,A_5\!\left(x,y\!=\!y_{*}\right) \ . 
\end{eqnarray} 
Then the KK decomposition of $A_\mu(x,y)$ is given by 
\begin{equation} 
A_\mu(x,y) =  
\sum_n A_\mu^{(n)}\!(x)\gaugemode{n}{y} \ , 
\end{equation}
where 
$\gaugemode{n}{y}$ is a linear combination of first order Bessel functions,
$J_1$ and $Y_1$,
\begin{eqnarray}
\gaugemode{n}{y}
= \frac{\sqrt{2k}\,e^{k |y|}}{N_n} 
\left[ 
J_1\!\left(\eigen{n}e^{k |y|}\right) +\alpha_n
Y_1\!\left(\eigen{n}e^{k |y|}\right) \right] \ ,  
\label{mode:gauge}
\end{eqnarray}
with $\eigen{n}\equiv{}\gaugemass{n}/k\neq{}0$.
The mass eigenvalues $M_n$ are determined by the condition 
$\partial_y \gaugemode{n}{\pi\rc}=0$, which leads to the equation 
\begin{equation}
J_0\!\left(\eigen{n}\right)
Y_0\!\left(\eigen{n}e^{\pi{}k\rc}\right)
=
Y_0\!\left(\eigen{n}\right)
J_0\!\left(\eigen{n}e^{\pi{}k\rc}\right) \ . 
\end{equation}
The masses $M_n$ grow linearly with $n$, being the first excited modes 
in the TeV range
($M_1 \sim 2.5 \, k e^{-\pi{}k\rc}$, $M_2 \sim 5.6 \, k e^{-\pi{}k\rc}$,
\ldots).

Continuity of $\partial_y \gaugemode{n}{y}$ at $y=0$ gives 
\begin{equation}
 \alpha_n
\,={}-\,\frac{J_0\!\left(\eigen{n}\right)}{Y_0\!\left(\eigen{n}\right)}
\end{equation}
and $N_n$ is a normalization constant which in the limit 
$M_n \ll k$ and $e^{\pikrc} \gg 1$ can be approximated by   
\begin{equation}
N_n \sim  e^{k \pi r_c} J_1\!\left(\eigen{n}e^{k \pi r_c}\right) \ .
\label{pr2}
\end{equation}

Several comments are in order. 
The wave function of the zero mode is $\chi_0 = 1/\sqrt{\pi r_c}$,
independent of the fifth coordinate, 
so it couples equally to both boundaries
with strength $g=g_{5D}/\sqrt{\pi r_c}$, being 
$g_{5D}$ the 5D gauge coupling. 
On the contrary, the excited modes are localized near the 
TeV boundary and have different couplings to fermions 
located on different branes, since these couplings are determined by 
the eigenfunctions $\gaugemode{n}{y}$ near the boundaries.
At the $y_*=\pi r_c$ boundary 
the term in (\ref{mode:gauge}) proportional to $J_1$ dominates
while at the other brane the eigenfunction can be well approximated
by the $Y_1$ term.
Within these approximations, the coupling of a gauge boson KK mode $n$ 
to 4D fermions is
(for $k r_c \simeq 12$):
\begin{eqnarray}
g^{(n)}/g \simeq &8.4 & \mbox{for the TeV boundary} \nonumber \\ 
g^{(n)}/g \simeq &.2/ \sqrt{n} & \mbox{for the $M_P$ boundary}
\label{gn} 
\end{eqnarray}  

The strong coupling of the KK gauge modes to fermions located on the 
TeV brane puts a restrictive constraint on this set-up.
In the limit where the KK tower exchanges can be described as 
a set of contact interactions, they lead to dimension six operators
which can be constrained by electroweak precision data, 
yielding a bound on the mass of the first gauge boson KK mode of order 
20 TeV \cite{KK1,KK3}.
Such a large scale may be a problem for the consistency of the theory, 
and seems to disfavor this scenario. However it is interesting to
consider it further, because the simplicity of the model allows 
to make definite calculations that illustrate generic features of
dynamical symmetry breaking in warped compactifications.

We have seen that the KK excitations of the SM gauge bosons couple strongly 
to the fermions located at the TeV brane and thus can produce bound states.
The analysis for bulk fermions is more involved, 
however it has been shown that the fermion zero-mode couples
strongly to the first KK gauge boson excitation, with 
$g^{(1)}/g \simeq 4.1$ (the coupling for higher $n$ is weaker)
\cite{KK3}.  
Although this result is obtained in the effective 4D theory, 
it seems to indicate that the coupling of bulk SM gauge bosons to the 
bulk fermion $\Psi$ may be also strong enough to form composite states 
involving the latter.

\section{Bound states}

In this section we study the formation of bound states in detail. 
The 5D gauge field theory is non renormalizable, and it 
remains weakly coupled beneath a local cut-off which depends on the 
position in the extra dimension, of order $\sim k e^{- k y}$ \cite{KK4}.
Moreover, at energies somewhat larger than this scale
the excited gravitons are strongly coupled and string/M-theoretic 
excitations should appear, which lie outside the domain of the 5D 
field theory.
Therefore near the TeV brane we expect the compositeness scale, 
$\Lambda$, to be given approximately by the cut-off of the 
effective 5D theory, not far above the scale of the first gauge boson 
KK excitations. 
Below the compositeness scale we integrate out the heavy gauge boson 
KK modes and approximate the resulting effective action 
by local four-fermion operators. These operators involve both, 
fermions $\psi(x)$ confined at the TeV brane and the bulk fermion 
$\Psi(x,y)$.

In order to construct the effective 5D theory which contains four-fermion 
operators we need the propagator for the bulk gauge bosons, given by  
\begin{equation}
\langle 0 | A_{\mu}(x', y')  A_{\nu}(x, y)|0  \rangle =
\int \frac{d^4 p}{(2 \pi)^4} \frac{2k}{N_n^2} e^{i p^{\mu} (x-x')_{\mu}} 
e^{(k |y|+k |y'|)}  \sum_n  
T_n(y') \frac{-i g^{\mu\nu}}{p^{\mu} p_{\mu}-M_n^2} T_n(y) 
\end{equation}
where we have defined
\beq 
T_n(y) \equiv  
 J_1\!\left(\eigen{n}e^{k |y|}\right) +\alpha_n
Y_1\!\left(\eigen{n}e^{k |y|}\right)  \ . 
\label{tn}
\eeq

Let us first consider the action corresponding to
the exchange of gauge boson KK excitations between fermions
confined at the TeV brane, $\psi(x)$, which reads
\begin{eqnarray} 
\mathcal{S}^{5D}&=&g_{5\D}^2 \int d^4 x \int d^4 x' \int dy \int
dy' \int \frac{d^4 q}{(2 \pi)^4} \sum_n e^{i q^{\mu} (x-x')_{\mu}}
\frac{2k}{N_n^2} \, e^{(k |y|+k |y'|)} T_n(y') \otimes \nonumber  
\\ && 
\frac{g^{\mu\nu}}{q^2-M_n^2} T_n(y)  
(\bar{\psi} \gamma_{\mu} T^r \psi)_{x',y'} \,
\delta(y'-\pi r_c)  \, (\bar{\psi} \gamma_{\nu} T^r \psi)_{x,y}  \,	 
\delta(y-\pi r_c)
\end{eqnarray}  
where the five dimensional gauge coupling is related to the effective 
four dimensional one $g$ by $g_{5\D}=g \sqrt{\pi r_c}$.

At scales $\Lambda$ above $M_1$ the action contains both 
light gauge boson KK modes and four-fermion operators obtained by 
integrating out the KK gauge bosons heavier than $\Lambda$, 
given by   
\begin{eqnarray} \mathcal{S}_{eff}^{5D}&=& -\sum_n g_{5\D}^2 
\frac{1}{M_n^2}\int
d^4 x \int d y \, 2 k \,e^{k |y|} \, \frac{T_n(y)}{N_n}  (\bar{\psi}
\gamma^{\mu} T^r \psi)_{x,y} \,\delta (y-\pi r_c)   \otimes \nonumber \\
&& \qquad  \left\{ \int dy' \,
e^{k |y'|} \, \frac{T_n(y')}{N_n} \, (\bar{\psi} \gamma_{\mu} T^r
\psi)_{x,y'}\, \delta(y'-\pi r_c) \right\} \ ,
 \label{bb1} \end{eqnarray}
with the sum over KK modes starting at $M_n \gsim \Lambda$.
The integration in $y'$ is trivial due to the delta function. Since we evaluate
$T_{n}(y)$ at the TeV brane, we can neglect the term proportional to $Y_1$
in (\ref{tn}). Using
the approximate form of the normalization constant $N_n$ given in eq.
(\ref{pr2}) and after Fierz transform, we obtain the familiar form of a
Nambu-Jona-Lasinio interaction 
\begin{eqnarray} \mathcal{L}_{eff}^{5D}
\supset \frac{c} {\Lambda^2}  \, \, \delta (y-\pi r_c)
(\bar{\psi}_{R}  \psi_{L}) (\bar{\psi}_{L} \psi_{R})  \ , 
\label{leff4} 
\end{eqnarray}  
where we have approximated the sum over KK modes by
\begin{eqnarray}
\sum_n \frac{1}{M_n^2} \sim \frac{1}{\Lambda^2} 
\ .
\end{eqnarray}

For the $SU(N)$ gauge groups, the coefficient $c$ is given by  
\beq
c = 3 k \pi r_c g_N^2
\left\{ C_2(\bar{\psi}_L) + C_2(\psi_R) - C_2(\bar{\psi}_L \psi_R) \right\}
\label{sun}
\eeq
where $g_N$ is the 4D coupling constant of $SU(N)$ and 
$C_2(r)$ the second Casimir invariant for the representation $r$
of the gauge group. 
For $U(1)_Y$, 
\beq
c = 3 k \pi r_c g_1^2
Y_{\bar{\psi}_L} Y_{\psi_R} 
\label{y}
\eeq
being $g_1$ the 4D $U(1)_Y$ coupling and $Y_r$ the hypercharge of the 
fermion $r$.

Obviously, four-fermion operators involving quarks  (the
left-handed $SU(2)_W$ doublet $\psi^3_L$  and the right-handed bottom $d^3_R$)
are more strongly coupled and  therefore more likely to form bound states.
The most attractive channels are scalars: an  $SU(2)_W$ doublet 
$\bar{\psi}^3_L d^3_R$ and a charged color triplet  
$\bar{\psi}^3_L \psi^{3c}_R$, with binding strength proportional to 
$3 k \pi r_c (\frac 4 3 g_3^2 - \frac {1}{18} g_1^2)$ and  
$3 k \pi r_c (\frac 2 3 g_3^2 + \frac 3 4 g_2^2 - 
\frac {1}{36} g_1^2)$,  respectively.

In \cite{MTY,BHL} it was shown that there is a critical value of the 
four-quark operator coefficient above which this attractive interaction
gives rise to a bound state.
In the large $N_c$ limit, the critical value is
$8 \pi^2 /N_c $.  For $k r_c \sim 11-12$, we easily obtain four-quark 
operator couplings larger than the critical one, thus there are bound states 
made of quarks located on the TeV brane. 
This is an expected result, because we have seen that the coupling of the 
excited modes of bulk gauge bosons to TeV brane fermions is stronger 
than the zero mode coupling.

Notice that the binding strength of the four-quark operators is 
supercritical for a wide range of the gauge couplings $g_i$. 
This is reassuring, because it means that the formation of bound 
states is   
not very sensitive to the running of the couplings above the TeV 
scale, which is still an open question in the Randall-Sundrum 
scenario \cite{KK4},\cite{ktz}.

If these bound states acquire vevs we have to face some 
phenomenological problems: $\langle \bar{\psi}^3_L \psi^{3c}_L \rangle$ 
would break charge and color, and  
since the Yukawa coupling of the doublet $\bar{\psi}^3_L d^3_R$ to
its constituents is typically large, the bottom quark would 
be too heavy, as it occurs in four dimensional top condensate models 
\cite{MTY,BHL,D1}. We will address these issues in the next section.

Let us consider now the action corresponding to
the exchange of gauge boson KK excitations among fermions
confined at the TeV brane and the bulk fermion $\Psi(x,y)$, 
\begin{eqnarray} 
\mathcal{S}^{5D}=g_{5\D}^2 \int d^4 x \int d^4 x' \int dy \int
dy' \int \frac{d^4 q}{(2 \pi)^4} \sum_n e^{i q^{\mu} (x-x')_{\mu}}
\frac{2k}{N_n^2} \, e^{(k |y|+k |y'|)} \otimes \nonumber  \qquad \\ 
\qquad T_n(y') \frac{g^{\mu\nu}}{q^2-M_n^2} T_n(y)  
(\bar{\Psi} \gamma_{\mu} T^r \Psi)_{x',y'} \,
  \, (\bar{\psi} \gamma_{\nu} T^r \psi)_{x,y}  \,	 
\delta(y-\pi r_c)\qquad\qquad
\end{eqnarray}  

In order to approximate this non-local interaction by four-fermion operators 
we use an interesting property of the Randall-Sundrum model: the warp factor 
produces a shift of a bulk field wave function depending on its 5D mass, 
so that massless fermionic fields are localized near the TeV brane, 
as can be seen from the 4D KK decomposition of fermion fields (\ref{fkk}).
Thus we can approximate the integral over the fifth coordinate $y'$ by 
a delta function  $\delta (y'-\pi r_c)$. 
This is just an approximation, but it is justified by the warp
factor, a feature not present in flat space scenarios. 

Analogously to the brane fermion case, the effective action at the scale 
$\Lambda$ is obtained by integrating out the gauge boson KK modes 
heavier than $\Lambda$,   
and after Fierz transform we get

\begin{eqnarray} 
\mathcal{L}_{eff}^{5D} \supset
\frac{c}{k \, \Lambda^2} \, \delta (y-\pi r_c)  \, 
(\bar{\Psi}_{R}  \psi^3_{L}) (\bar{\psi}^3_{L} \Psi_{R})  \ ,
\label{leff5} 
\end{eqnarray} 
with $c$ as defined in eqs. (\ref{sun}) and (\ref{y}).
The factor $1/k$ appears for dimensional reasons, but it will be 
compensated by the normalization constant of the bulk fermion field,
eqs. (\ref{fer1})-(\ref{zeromode}).

The light bound states would be four-dimensional scalars, namely 
$ H \propto \bar{\psi}_L^3 \Psi_R$, 
which we will identify with the Higgs doublet, and 
$\bar{\Psi}_R d^{3c}_L$. The binding strength of the corresponding 
four-quark operators is 
$3 k \pi r_c (\frac 4 3 g_3^2 + \frac {1} {9} g_1^2)$ and 
$3 k \pi r_c (\frac 2 3 g_3^2 + \frac {2} {9} g_1^2)$, 
so $H$ is more deeply bound than the $SU(2)_W$ singlet
$\bar{\Psi}_R d^{3c}_L$. 
The $\Psi$ field in eq.(\ref{leff5}) is five dimensional 
(although the delta function forces $y=\pi r_c$), 
so it is not possible
to apply the result of \cite{BHL} on the critical coupling in a 
straightforward manner, because it was obtained by solving the gap
equation in four dimensions. However, since the bulk fermion  
is localized in the region where the first KK
mode of the gauge bosons couples strongly to the fermion zero-mode, 
we expect that these bound states are also produced.
We assume that they are, and proceed with the analysis of the 
phenomenological implications of this scenario.

Finally, there is also the possibility of forming a five-dimensional 
gauge singlet composite scalar, $\bar{\Psi} \Psi$. 
However since its wave function vanishes at the TeV boundary, there is 
no quartic coupling involving this singlet and the previously 
discussed four-dimensional bound states, and thus it has no effect 
in the low energy effective theory.

Therefore in the Randall-Sundrum scenario, with only 
one extra dimension, strongly coupled four-quark operators are naturally 
induced and give rise to 4D bound states. 
The next step is to compute the parameters of the composite fields 
produced by this condensation of quark pairs.

\section{Effective theory in 5D}
 
To derive the low energy effective Lagrangian we follow the procedure 
of ref.\cite{BHL}.  
We use the auxiliary field method to construct scalar fields from a pair of 
quarks and we assume that they become propagating degrees of freedom below 
the compositeness scale, $\Lambda$.
We consider just the most deeply-bound channels, which lead to the 
lightest scalars, more likely to get negative squared-masses and 
acquire vevs.
Thus  we define 
\begin{equation}
H = -\frac{1}{\Lambda^2} \sqrt{\frac{c_H}{k}} ( \bar{\psi}^3_{L} \Psi_{R}) 
\end{equation}
and
\begin{equation}
\phi = -\frac{1}{\Lambda^2} \sqrt{c_{\phi}} (\bar{\psi}^3_{L} d^3_{R})  \ ,
\end{equation}
where $c_H = 3 \pi k r_c (\frac 4 3 g_3^2 + \frac {1} {9} g_1^2)$ and 
$c_\phi = 3 \pi k r_c (\frac 4 3 g_3^2 - \frac {1} {18} g_1^2)$.
Introducing these definitions in (\ref{leff4}) and (\ref{leff5})
we obtain the following effective Lagrangian at the compositeness scale
\begin{eqnarray} \mathcal{L}^{5D}[\Lambda]=-\delta (y-\pi r_c) 
\left\{ \sqrt{\frac{c_H}{k}}
(\bar{\Psi}_{R}  \psi^3_{L}) H + \Lambda^2 H^{\dagger} H- 
 \sqrt{c_{\phi}}
(\bar{d}^3_R  \psi^3_{L}) \phi + \Lambda^2 \phi^{\dagger} \phi \right\} 
\ .
\end{eqnarray} 
At scales $\mu < \Lambda$, the Yukawa interactions will induce kinetic terms 
for the scalars, as well as an effective potential which includes 
mass and quartic terms:
\begin{eqnarray}
 \mathcal{L}^{5D}[\mu]=\delta(y-\pi r_c)\left\{ 
Z_H(\mu) D^{\nu} H^{\dagger} D_{\nu} H+
 Z_{\phi}(\mu) \partial^{\nu} \phi^{\dagger} \partial_{\nu} \phi  
 - \sqrt{\frac{c_{H}}{k}}  H
\bar{\psi}^3_L  \Psi_{R} - \sqrt{c_{\phi}}
(\bar{d}^3_{R}  \psi^3_{L}) \phi 
 - V(\mu) \right\}
\label{leff}
\end{eqnarray}
where 
\begin{eqnarray} 
V(\mu)=\delta(y-\pi r_c)\left\{ M_H^2 H^{\dagger} H +
\frac{\lambda_H}{2} (H^{\dagger} H)^2 + m^2_{\phi} \phi^{\dagger} \phi+
\frac{\lambda_{\phi}}{2} (\phi^{\dagger} \phi)^2 + 
\lambda_{\phi H} H^{\dagger} H
\phi^{\dagger} \phi\right\}
\end{eqnarray}
We compute the parameters of the effective Lagrangian
$\mathcal{L}^{5D}[\mu]$ in the large $N_c$ limit, where only one fermion
loop contributes. We thus need the bulk fermion propagator, given by
\begin{eqnarray} \langle 0 | \Psi(x', y')  \bar{\Psi}(x, y)|0  \rangle &=& 
\int \frac{d^4 p}{(2 \pi)^4} e^{i p^{\mu} (x-x')_{\mu}} \frac{2k}{1-e^{-\pi k
r_c}} \, e^{-\frac{k}{2} (|\pi r_c-y|+|\pi r_c - y'|)} \, e^{-\frac{3}{2} (k
|y|+ k|y'|)} \, \frac{i}{1+\delta_{n0}} \otimes \nonumber \\  && \sum_n\, (c_n(y') P_R+s_n(y')
P_L) \, \frac{\gamma^{\mu} p_{\mu}+\gamma_5 m_n}{p^{\mu} p_{\mu}-m_n^2} \,
(s_n(y) P_R+c_n(y) P_L) \end{eqnarray}
where $c_n(y) =\cos \left\{ \frac{m_n}{k}\left(e^{k |y|}-1\right) \right\}$ 
and $s_n(y)$ is defined analogously.
At leading order in the $1/N_c$ expansion the mass parameters are
\begin{eqnarray} 
M_H^2&=&\Lambda^2- N_c c_{H}  \frac{4}{1-e^{-\pi k
r_c}} \sum_n \int \frac{d^4 k}{(2 \pi)^4} \frac{-i }{k_{\nu} k^{\nu}-m_n^2} 
\\
m_{\phi}^2&=&\Lambda^2-4 N_c c_{\phi} \int \frac{d^4 k}{(2 \pi)^4} \frac{-i
}{k_{\nu} k^{\nu}} 
\end{eqnarray}
The above integrals are defined with a cutoff at the compositeness scale 
$\Lambda$. Assuming that the number of KK modes at the scale $\Lambda$,  
$n_{KK}(\Lambda)=(e^{\pi k r_c}-1) \Lambda / \pi k$, is large enough, we
can approximate the sum over KK modes by an integral, 
$\sum_n \rightarrow \int dn$. 
Using that 
\begin{eqnarray}
\frac{-i}{8 \pi^2} \int_0^{n_{KK}} dn \int_0^1 dp 
\frac{p}{p^2+(m_n/\Lambda)^2}=
-\frac{in_{KK}}{16 \pi^2} f_1
\end{eqnarray}
where  
\begin{equation}
f_1=\int_0^1 dq \int_0^1 p^2 dp^2 \frac{1}{p^2+q^2} \sim 0.63 \ .
\end{equation}
we obtain
\begin{eqnarray}
M_H^2&=&\Lambda^2
\left(1-n_{KK}(\Lambda) N_c \frac{c_{H} }{4 \pi^2} 
\frac{f_1}{1-e^{-\pi k r_c}} \right) 
\label{M}\\ 
m_{\phi}^2&=&\Lambda^2
\left( 1- N_c \frac{c_{\phi}}{4 \pi^2} \right) 
\label{mp}
\end{eqnarray}
From eqs. (\ref{M}),(\ref{mp})
we see that the Higgs mass 
$M_H^2$ can be negative while $m^2_{\phi}$ remains positive, because the 
contribution of the bulk fermion KK modes in the loop decreases
$M^2_H$. 
Since the compositeness scale $\Lambda$ is above the electroweak 
scale, $c_H$ should be close to the critical value for which 
$M_H^2$ becomes negative. In this region, $H$ acquires a vev and 
provides a positive contribution to $m^2_{\phi}$ through their coupling,
$\lambda_{\phi H}$. The resulting mass $m^2_{\phi}$ is then expected to stay
positive and large. 
By the same argument, other bound states less deeply-bound than $H$ and 
$\phi$ will also have positive squared-masses, preventing charge and
color breaking, as well as a too heavy bottom quark. 
Therefore, naturally we obtain that the only composite field which acquires
a vev is $H \propto \bar{\psi}_L^3 \Psi_R$, due to the KK excitations of
one of its constituents.   
These excitations will give also a natural framework for a light top mass, 
as we will see in the next section.

The wave function renormalization and self-coupling of the Higgs
in the large $N_c$ limit are given by 
\begin{eqnarray}
Z_H&=&N_c c_{H}   \frac{2 }{1-e^{-\pi k r_c}} \sum_n \int 
\frac{d^4 k}{(2 \pi)^4} \frac{-i }{k_{\mu} k^{\mu} (k_{\nu} k^{\nu}-m_n^2)} 
\\
\lambda_H&=&8 N_c c_{H}^2  \left( \frac{1}{1-e^{-\pi k r_c}} \right)^2 
\sum_{n_1,n_2} \int \frac{d^4 k}{(2 \pi)^4} (-1)^{n_1+n_2}
\frac{-i }{(k_{\nu} k^{\nu}-m_{n_1}^2)(k_{\nu} k^{\nu}-m_{n_2}^2)}  \ .
\end{eqnarray}
Again, we approximate the sums over KK modes by integrals and we find
\begin{eqnarray}  
Z_H&=&N_c c_{H}  \frac{1}{1-e^{-\pi k r_c}}
\frac{n_{KK}(\Lambda)}{8 \pi^2} f_2 \\  
\lambda_H&=&n_{KK}(\Lambda)^2 \frac{N_c }{2
\pi^2} \left(\frac{ c_{H} }{1-e^{-\pi k r_c}}\right)^2 f_3   
\end{eqnarray}
where
\footnote{Note that these integrals are the same of \cite{D1} with 
$f_1= F_3(L), f_2=F_1(L), f_3=F_5(L)$, where $L \sim (TeV)^{-1}$ is the  
length of the fifth dimension}  
\begin{eqnarray} f_2&=&\int_0^1 dq \int_0^1  dp^2
\frac{1}{p^2+q^2} \sim 2.26  \\ 
f_3&=&\int_0^1 dq \int_0^1  dq' \int_0^1 p^2 dp^2
\frac{1}{(p^2+q^2)(p^2+q'^2)} \sim 2.71  \ .  
\end{eqnarray}

 Redefining $H \rightarrow \sqrt{Z_H} H$ to obtain a canonical kinetic term, 
the Higgs mass becomes 

\begin{eqnarray} 
\bar{M}_H^2=\frac{M_H^2}{Z_H}=\frac{2 \Lambda^2}{f_2}\left( \frac{4
\pi^2 (1-e^{-\pi k r_c})}{ N_c c_{H}   n_{KK}}- f_1 \right)
\label{masaHiggs} 
\end{eqnarray}
and the quartic coupling

\begin{eqnarray} \bar{\lambda}_H=\frac{\lambda_H}{Z_H^2}=\frac{32 \pi^2}{N_c}
\frac{f_3}{f_2^2} \sim 60 \ .
\label{acoplamiento} 
\end{eqnarray}

In this scenario we have a heavy Higgs boson with a large (non-perturbative)
quartic coupling, which is generic from top mode scenarios. Having a heavy
Higgs boson is not in contradiction with data, nor with triviality bounds 
because the cut-off of the effective theory, the compositeness scale, 
is very low. 
Regarding the unitarity bound from the longitudinal WW scattering cross 
section, the Higgs mass we obtain at tree level, 
$m_H = v / \sqrt{\bar{\lambda}_H}$, is above this bound.
However, with such a large quartic self coupling we cannot trust the tree 
level relation between the mass and the vev of the Higgs, 
and we can only take it as a rough estimate of the Higgs mass.

Next-to-leading order corrections in the $N_c$ expansion (i.e., 
contributions from gauge bosons and composite scalar loops) could be
included by evolving the couplings from the compositeness scale 
$\Lambda$ down to the electroweak scale. To do so, we would need the
$\beta$-functions for the four-dimensional SM couplings in the 
Randall-Sundrum scenario, which have not been computed yet.
However we do not expect that next-to-leading effects would 
change qualitatively our results.

\section{Fermion Masses}
\label{fm}

In this section we discuss the generation of fermion masses within the 
simplest set-up described in sec. \ref{setup}. First, we calculate the 
top quark mass and then we include the other two generations in the 
model.

\subsection{Top  Mass}

To compute the top quark mass we have to canonically normalize the Higgs 
kinetic term in the effective 5D Lagrangian (\ref{leff}) 
and integrate $\mathcal{L}_{eff}^{5D}$ over the fifth dimension 
$y$ to obtain the 4D effective Lagrangian. 
We find 
\begin{eqnarray}
\mathcal{L}^{4D} \supset -y_t  H \bar{\Psi}_{R}(x,\pi r_c) \psi^3_{L}(x) 
\end{eqnarray}
where the top Yukawa coupling $y_t$ is given by
\begin{eqnarray}
y_t = \sqrt{\frac{c_{H}}{Z_H (1-e^{-\pi k r_c})}} = 
\frac{2\sqrt{2}\pi}{\sqrt{N_c n_{KK} f_1}} \ .
\end{eqnarray}
Recall that $n_{KK}$ is the number of KK modes produced at the 
compositeness scale $\Lambda$.  
Thus we obtain that the top Yukawa coupling is suppressed by
the factor $\sqrt{n_{KK}}$, and turns out to be order one for 
$n_{KK} \sim 10$, i.e., $\Lambda \sim 10 \, M_1$, with $M_1$ 
the mass of the first gauge boson KK excitation.
This result is also generic in top condensate models with flat 
extra dimensions \cite{D1,D2}, avoiding a too heavy top quark 
typical of four dimensional top condensate models.

The zero mode of $\Psi_R(x,\pi r_c)$ becomes the four-dimensional
right-handed top quark, $t_R$. 
Taking into account the KK decomposition of the bulk fermion (\ref{fkk}),
when the Higgs acquires a vev 
$\langle H^0 \rangle = v/\sqrt{2}$ we obtain the following mass 
matrix for the $t_L$ component of $\psi_L$ and the $\Psi$ KK modes:
\begin{equation}
\left (
\ba{cccc}
\bar{t}_L  & \bar{\Psi}_{L}^{(1)} &   \bar{\Psi}_{L}^{(2)} & \ldots             \ea \right)
\left (
\ba{ccccc}
y_t \frac{v}{\sqrt{2}} & y_t v &  y_t v & y_t v & \ldots \\ 
0  & \frac{\pi k}{e^{\pi k r_c}-1}  & 0  & 0  & \ldots  \\
0  & 0  &\frac{2 \pi k}{e^{\pi k r_c}-1}   & 0  & \ldots \\
\ldots &\ldots &\ldots &\ldots &\ldots 
\ea \right) 
\left (
\ba{c}
t_R \\ 
\Psi_{R}^{(1)} \\
\Psi_{R}^{(2)}\\
\ldots  \ea \right) 
\end{equation}
The diagonal entries are the usual KK mass terms.
The top mass is given by the lowest eigenvalue of this matrix.
To compute it, 
we can perturbatively diagonalize the mass matrix,  
using that  $m_n=n \pi k e^{-\pi k r_c} > v $ because KK excitations have not 
been observed at the moment. 
Thus, an expansion in 
$\left( \frac{v}{\pi  ke^{-\pi k r_c} } \right)^2 \ll 1$ is justified and  
within this approximation the lowest eigenvalue is
\begin{equation}
m_t = \frac{y_t v}{\sqrt{2}} \left[ 1-{\cal O}\left(
\frac{ v }{\pi k e^{-\pi k r_c} }\right)^2 \right] \ ,
\label{masatop}
\end{equation}
which gives easily a top mass in the experimental range $\sim 170$ GeV for 
$n_{KK} \sim 10$.

\subsection{Flavor Symmetry Breaking}

Once we have accommodated the electroweak symmetry breaking scale and the 
top mass, the next step is to generate masses for the remaining 
quarks and leptons of the Standard Model. 
The simplest possibility is that the fermions of the first two generations 
are confined at the TeV brane. Then, there would be three degenerate 
Higgs doublets
$H^i \propto \bar{\psi}^i_{L} \Psi_{R} \ ,(i=1,2,3)$, which obtain a vev and 
break the $U(2)_\psi \times U(2)_u$ flavor symmetry, leading to two 
Nambu-Goldstone bosons, besides the one eaten by the electroweak gauge
bosons. Obviously, it is necessary a source of flavor symmetry breaking. 
When fermions are located near the TeV brane, higher-dimensional operators
are suppressed by the multi-TeV scale, so generically there would be 
four-quark operators of the type \cite{D1,D2}
\beq
\frac{\eta_{i}}{\Lambda^2} 
(\bar{\psi}^i_{L} \Psi_{R}) (\bar{\Psi}_{R} \psi^i_{L})  \ ,
\eeq
where the coefficients $\eta_i$ are ${\cal O}(1)$ and can be treated
perturbatively. We have seen that the squared-mass of a composite field 
depends on the strength of the interaction between its constituents 
(see eq.(\ref{M})), thus assuming that the coefficient $\eta_3$ is 
slightly larger than the others, it is possible that only $H^3$ 
gets a vev, giving a mass of the electroweak scale just to the 
top quark. 
The rest of bound states may be quite heavy, even with small
differences among the flavor symmetry breaking coefficients
$\eta_i$.

Of course, if all higher-dimensional operators consistent with the SM 
symmetry are only suppressed by the multi-TeV scale, flavor-changing 
effects and proton decay become a problem, much as in the 
original Randall-Sundrum set-up.
We do not attempt to solve these problems here, and we just assume that 
dangerous flavor-changing and baryon number violating operators 
are suppressed by some mechanism of the underlying 
fundamental theory.

 If the only light composite scalar is $H^3$, to produce the observed 
pattern of fermion masses and mixings we shall consider the presence of 
the effective operators 
\begin{eqnarray} 
\frac{1}{\Lambda^2} ( \bar{\Psi}_R  \psi^3_{L}) 
\left( \lambda^u_{ij} \bar{\psi}^i_L u^j_R
+  \lambda^d_{ij} \bar{\psi}^i_L i \sigma_2 d^j_R +  
\lambda^e_{ij}
\bar{l}^i_L i \sigma_2 e^j_R \right) \ ,
\label{masses}
\end{eqnarray} 
which would lead to Yukawa couplings of the $H^3$ doublet to the 
SM fermions. As usual, large hierarchies in the coefficients
$\lambda_{ij}$ 
should be assumed in order to explain the observed fermion masses.

\section{The Standard Model in the Bulk}
\label{other}
The minimal model studied in the previous sections should be regarded 
as a `working proof' of dynamical symmetry breaking in the Randall-Sundrum
scenario, but it is not by any means unique. 
A more natural possibility is that all SM fermions propagate 
in the 5D bulk. 
In this case, the approximation of the 
higher-dimensional gauge interactions by local four-fermion operators
is questionable and the issue of whether dynamical symmetry breaking 
really takes place or not is highly non trivial \cite{hty}. 
A careful study of such dynamical problem is beyond the scope 
of this paper, but given the strong coupling between gauge 
boson KK modes and the zero mode of fermions localized near the 
TeV brane, we argue that the condensation 
of 5D massless quarks seems very likely.

The localization of fermion fields living in a slice of $AdS_5$ depends 
on its 5D masses \cite{gp}.
Left-handed (right-handed) zero modes of 
fermions with bulk mass terms $M_{5D}/k < 1/2$ ($M_{5D}/k > - 1/2$)
(in particular massless fermions, as we have seen) 
live near the TeV brane, while the left-handed (right-handed) zero modes 
of fermions with $M_{5D}/k > 1/2$ ($M_{5D}/k < - 1/2$)
are localized near the Planck boundary. 
For left-handed (right-handed) zero modes 
$M_{5D}/k = 1/2$ ($M_{5D}/k = - 1/2$)
corresponds to the conformal limit, and in this case the zero mode is flat. 
On the other hand, KK modes of gauge bosons are always located near
the TeV brane and therefore they will couple more strongly to 
light fermions
\footnote {In particular, at the conformal limit KK gauge bosons do 
not couple to the fermion zero mode, because the five-momentum is 
conserved in this limit.}. 
We thus assume that the left-handed third generation quarks 
$\Psi^3$ have $M_{5D}/k \lsim 1/2$
and the right-handed top quark $U^3$ has 
$M_{5D}/k \gsim - 1/2$, and due to the strong coupling to gauge 
boson KK modes they form bound states. 
On the contrary, if the remaining SM left-handed (right-handed)
fermions have 5D mass terms $M_{5D}/k \gsim 1/2$
($M_{5D}/k \lsim - 1/2$),
their zero modes live near the Planck 
brane, where the coupling to gauge boson KK modes is too weak 
to produce composite states.

Under these assumptions, we expect that the SM gauge interactions in the 
bulk will induce only condensates of third generation quarks 
$\Psi^3$ and $U^3$. 
Using the most attractive channel analysis \cite{D2} one finds that 
the channel $H = \bar{\Psi}^3 U^3$ is the most attractive one among 
those which transform non-trivially under the SM group, but 
there would be gauge-singlet scalars more tightly bound than $H$,
namely $\bar{\Psi}^3 \Psi^3$ and $\bar{U}^3 U^3$.
As a consequence, these singlets would acquire vevs which do not 
break any gauge symmetry and provide a positive 
contribution to the squared-mass of the $SU(2)_W$ doublet $H$. 
In this case, it is not clear whether the scalar effective potential 
is minimized by a nonzero vev of $H$; moreover, simple estimates are 
not reliable due to the non-perturbative nature of 
the quark condensation. 
To avoid this problem we could incorporate more $Z_2$ symmetries during 
the orbifold projection, preventing the existence of singlet zero 
modes in the KK decomposition \cite{D2}. 
Thus the lightest modes of the singlets would have squared masses less 
negative, allowing electroweak symmetry breaking.

If the composite Higgs $H$ acquires a vev, it will provide 
electroweak scale masses for the gauge bosons and the top quark.
It is important to note that since the composite Higgs has a 5D mass of 
order TeV, it is also localized near the TeV boundary. 
In order to explain the SM fermionic spectrum we have to assume the 
presence of four-fermion operators of the type (\ref{masses}).  
However now the 5D Yukawa couplings can be all of order 
$\lambda^{(5)}_{ij} \, k \sim 1$, because the hierarchy of masses is
originated by the exponentially small overlap of the fermions
localized near the Planck brane with the composite Higgs field, much 
as in the model of ref.\cite{gp} with a fundamental SM Higgs localized 
on the TeV brane.
There, it is shown that the spectrum of the SM fermionic sector can be 
naturally generated just assuming small splittings in the bulk fermion 
masses, all in the range $|M_{5D}|/k \sim 1$. 
We refer the reader to \cite{gp} for details. 
This idea of using the warp factor to explain fermion mass 
hierarchies has also been consider in \cite{KK2} in the context of
neutrino masses.

Finally, we should point out that in this scenario FCNC 
dangerous operators
are safely suppressed, but proton decay is still a problem 
\cite{gp}.
 
\section{Conclusions}
We have studied dynamical electroweak symmetry breaking in the Randall-Sundrum
scenario and we have shown that it is possible to break the electroweak 
symmetry using the ideas of top condensate mode in the context of one
warped extra dimension, with no fundamental fields beyond the 
SM gauge bosons and fermions living in the 5D bulk.

In warped compactifications, the coupling of gauge boson KK excitations 
to fermions depends on the position of the fermion field 
in the bulk. In particular, there is a sizeable enhancement of their
couplings to fermions localized near the TeV brane, as compared with the
gauge boson zero mode coupling. Such a strong interaction 
produces quark pair condensates, some of which can acquire vevs and  break 
dynamically the electroweak symmetry. 
Contrary to what happens in flat extra dimensions \cite{D1}, we 
do not have to rely on the degeneracy of KK modes in $D \ge 6$ 
dimensions to trigger dynamical symmetry breaking.

We have considered a minimal model, where only gauge bosons and the 
right-handed top quark propagate in the 5D bulk, while all the other 
SM fermions are confined on the TeV brane. 
The Higgs boson emerges as a bound state of the left-handed third 
generation quarks and the right-handed top.
The simplicity of the model allows to perform reliable calculations 
within the four-fermion operator approximation. 
We have computed, at leading order in the $N_c$ expansion, the parameters 
of the composite scalar effective theory and the top quark mass.
The reason why at least the right-handed top quark must live in the 
bulk is twofold. 
First, its KK excitations active at the compositeness scale give
a large and negative contribution to the squared-mass of the
$SU(2)_W$ doublet composite Higgs, which then acquires a
vev and breaks the electroweak symmetry. 
Second, the number of its active KK excitations 
suppresses the top quark Yukawa coupling, leading to a top mass
within the experimental range. 
Although our calculation only includes the 
leading $N_c$ contribution, it shows that dynamical 
electroweak symmetry breaking is feasible in the Randall-Sundrum model, 
and we do not expect that next-to-leading order corrections will 
invalidate this conclusion.

We have also considered a more natural scenario, in which all SM fermions 
as well as the gauge bosons propagate in the bulk. In this case, 
a local four-fermion approximation of the strong interactions mediated by
KK gauge boson modes is not justified, and we can only describe 
qualitatively the expected features of the model.
Interestingly enough, just assuming that the light fermions are localized 
near the Planck brane and the heavy ones near the TeV boundary, we argue 
that only composite states bound out of third generation 
quarks are produced and we can explain the fermion mass hierarchies. 
The question of whether the composite Higgs doublet 
does acquire a nonzero vev requires further investigation.

\section*{Acknowledgments}

V.S. would like to thank Bogdan Dobrescu for interesting suggestions and
comments, and Lisa Randall for enlightening discussions.
We also thank Pilar Hern\'andez and Arcadi Santamar\'\i a for comments on  
the manuscript. V.S. would like to thank the Center for Theoretical Physics 
of the Massachusetts Institute of Technology for its hospitality while this 
work was initiated.
This work was supported in part by the spanish DGESIC grants 
PB97-1261 and PB98-0693, by the Generalitat Valenciana 
under grant GV99-3-1-01 and by the TMR network 
contract HPRN-CT-2000-00148 of the European Union.


\begin{thebibliography}{99}

\bibitem{XL}

N.~Arkani-Hamed, S.~Dimopoulos and G.~Dvali,
Phys.\ Lett.\  {\bf B429} (1998) 263
[hep-ph/9803315].
I.~Antoniadis, N.~Arkani-Hamed, S.~Dimopoulos and G.~Dvali,
Phys.\ Lett.\  {\bf B436} (1998) 257
[hep-ph/9804398].


\bibitem{RS}
L.~Randall and R.~Sundrum,
Phys.\ Rev.\ Lett.\  {\bf 83} (1999) 3370
[hep-ph/9905221].

\bibitem{MTY}
V.A. Miransky, M. Tanabashi and K. Yamawaki,
Phys. Lett. {\bf B221} (1989) 177;
Mod. Phys. Lett. {\bf A4} (1989) 1043.

\bibitem{BHL}
W.~A.~Bardeen, C.~T.~Hill and M.~Lindner,
Phys.\ Rev.\  {\bf D41} (1990) 1647.

\bibitem{seesaw}
B.~A.~Dobrescu and C.~T.~Hill,
Phys.\ Rev.\ Lett.\  {\bf 81} (1998) 2634
[hep-ph/9712319].

\bibitem{AD}
N.~Arkani-Hamed and S.~Dimopoulos, [hep-ph/9811353].


\bibitem{K}
A.~B.~Kobakhidze, [hep-ph/9904203].

\bibitem{D1}
B.~A.~Dobrescu,
Phys.\ Lett.\  {\bf B461} (1999) 99 [hep-ph/9812349].
H.~Cheng, B.~A.~Dobrescu and C.~T.~Hill,
Nucl.\ Phys.\  {\bf B589} (2000) 249
[hep-ph/9912343].


\bibitem{D2}
N.~Arkani-Hamed, H.~Cheng, B.~A.~Dobrescu and L.~J.~Hall,
Phys.\ Rev.\  {\bf D62} (2000) 096006
[hep-ph/0006238].



\bibitem{KK0}
W.~D.~Goldberger and M.~B.~Wise,
Phys.\ Rev.\ D {\bf 60} (1999) 107505
[hep-ph/9907218].


\bibitem{KK1}
H.~Davoudiasl, J.~L.~Hewett and T.~G.~Rizzo,
Phys.\ Lett.\  {\bf B473} (2000) 43 [hep-ph/9911262].
A.~Pomarol, Phys.\ Lett.\  {\bf B486} (2000) 153
[hep-ph/9911294].


\bibitem{KK2}
Y.~Grossman and M.~Neubert,
Phys.\ Lett.\  {\bf B474} (2000) 361 [hep-ph/9912408].

\bibitem{KK3}
S.~Chang, J.~Hisano, H.~Nakano, N.~Okada and M.~Yamaguchi,
Phys.\ Rev.\  {\bf D62} (2000) 084025
[hep-ph/9912498].

\bibitem{gp}
T.~Gherghetta and A.~Pomarol,
Nucl.\ Phys.\  {\bf B586} (2000) 141
[hep-ph/0003129].


\bibitem{KK4}
A.~Pomarol,
Phys.\ Rev.\ Lett.\  {\bf 85} (2000) 4004
[hep-ph/0005293].

\bibitem{DHR}
H.~Davoudiasl, J.~L.~Hewett and T.~G.~Rizzo,
Phys. Rev. {\bf D63} (2001) 075004
[hep-ph/0006041].


\bibitem{AS}
F. del Aguila and J.~Santiago, Phys.\ Lett.\ {\bf B493} (2000) 175 
[hep-ph/0008143].


\bibitem{ktz}
J.~ Kubo, H.~Terao and G.~Zoupanos, [hep-ph/0010069].


\bibitem{hty}
M.~Hashimoto, M.~Tanabashi and K.~Yamawaki,
[hep-ph/0010260].

 
\end{thebibliography}
\end{document}